\documentclass[pra,amsmath,amsfonts,superscriptaddress]{revtex4}
\usepackage[bookmarksnumbered, colorlinks,plainpages]{hyperref}
\hypersetup{colorlinks,linkcolor=blue,citecolor=blue,urlcolor=blue}
\usepackage{epsf,epsfig}
\usepackage{amssymb}
\usepackage{stmaryrd}
\usepackage{amsmath}
\usepackage{MnSymbol}
\usepackage{float}
\usepackage[caption=false]{subfig}
\usepackage{color,soul}
\usepackage{xcolor}
\usepackage{pgf,tikz}
\usepackage{graphicx}
\usepackage{dcolumn}
\usepackage{bm,bbm}
\usepackage{color,soul}
\usepackage{xcolor}
\usepackage{import}
\usepackage{csquotes}
\usepackage{amsthm}
\usepackage[toc,page]{appendix}
\graphicspath{{./fig/}}
\usepackage{enumerate}
\usepackage{listings}
\usepackage{soul}
\usepackage[]{xcolor}
\usepackage{chngcntr}

\counterwithin*{example}{section}
\newcommand{\eref}[1]{Eq.~(\ref{#1})}

\newcommand{\Tr}{{\mathrm{Tr}}}

\DeclareMathOperator{\diag}{diag}

\newcommand{\ket}[1]{|#1\rangle}

\newcommand{\e}{{\mathrm e}}

\def\a{\alpha}

\def\g{\gamma}

\def\be{\begin{equation}}
\def\ee{\end{equation}}
\def\ba{\begin{eqnarray}}
\def\ea{\end{eqnarray}}
\def\la{\langle}
\def\ra{\rangle}
\def\a{\alpha}

\def\h{\hskip 1cm}

\def\lo{\longrightarrow}

\begin{document}
\title{ Simulating of X-states and the two-qubit XYZ Heisenberg system  on IBM quantum computer}

\author{Fereshte Shahbeigi\footnote{email: fereshteshahbeigi@gmail.com}}
\affiliation{\it Department of Physics, Sharif University of Technology, Tehran 14588, Iran}
\affiliation{\it Center for Theoretical Physics, Polish Academy of Sciences, Warsaw, Poland}
\author{Mahsa Karimi\footnote{email: m.karimii1997@gmail.com }}
\affiliation{\it Department of Physics, Sharif University of Technology, Tehran 14588, Iran}
\author{Vahid Karimipour\footnote{email: vahid.karimipour@gmail.com}}
\vspace{1cm}
\affiliation{\it Department of Physics, Sharif University of Technology, Tehran 14588, Iran}

\date{Jan. 6, 2022}

\begin{abstract}
Two qubit density matrices which are of X-shape, are a natural generalization of Bell Diagonal States (BDSs) recently simulated on the  IBM quantum device.  We generalize the previous results and propose a quantum circuit for simulation of a general two qubit X-state, implement it on the same quantum device, and study its entanglement for several values of the extended parameter space. We also show  that their  X-shape is approximately robust against noisy quantum gates.  To further physically motivate this study, we invoke the  two-spin Heisenberg XYZ system and show that for a wide class of initial states, it leads to dynamical density matrices which are X-states. Due to the symmetries of this Hamiltonian, we show that by only two qubits, one can simulate the dynamics of this  system on the IBM quantum computer.
\end{abstract}
\maketitle
\section{Introduction}
\noindent The significant technological improvements in the last decade
 have brought us closer than ever to the dream of having a quantum computer
 satisfying some demanded criteria \cite{D00}.
By a quantum computer, one anticipates  accomplishing tasks that cannot be done
on classical computers in a reasonable time scale, and simulation of quantum systems is one of the most promising of these tasks \cite{N00}.
In this regard one of the most impressive advances is the IBM Quantum Experience \cite{ibm} which makes a limited number of qubits public and ready to be programmed remotely. The accompanying open source Qiskit software \cite{qiskit} allows to compare simulation of any algorithm run on this software and the actual IBM quantum device,  hence assessing the level of noise and gate imperfections and their effect on the output of any algorithm. Quite recently this has led many groups to run interesting algorithms on this quantum computer with promising results \cite{Ns20,Al18,Ay17,Ab20,G19,PM19,Di18,ABH20,MBP21,BCZ21}.  Among these works, the ones which have inspired our work are the study of a class of two-qubit states, namely Bell diagonal and Werner states, specially with regard to their correlation properties \cite{G19,PM19}. These are classical mixtures of the form,
\be\label{bds}
\rho_{_{BDS}}=\sum_{i=1}^4
p_i |\phi_i\ra\la \phi_i|\ee
where $\sum_i p_i=1$, and $|\phi_i\ra$'s are the maximally entangled Bell states. These states are characterized by three independent real parameters, namely the three independent parameters which characterize the probability distribution $\{p_i\}$. In \cite{G19}, a complete study of correlation properties , i.e. entanglement \cite{RFW,B96,W98}, discord \cite{OL01,M10,M12}, non-locality \cite{JSBELL,H95} and steering \cite{Wi07,J07} of these states were performed by first generating them on a circuit and then making appropriate measurements. The results of Qiskit simulations 
were in pretty good agreement with those of the actual IBM quantum computer. Having three real parameters residing inside a tetrahedron, these  results could nicely be depicted on various graphs of tetrahedrons. \\

\noindent Inspired by this work, we ask if one can extend this study to a larger class of states and if such an extension gives us new insight about the performance of the IBM quantum computer with its limited number of publicly available qubits. To this end we note that the Bell diagonal states are very special  in a larger class of states, aptly called X states \cite{YE07}, due to their shape,
\begin{equation}
\label{x shape}
{\rho _X} = \left( {\begin{array}{*{20}{c}}
	a&0&0&w\\
	0&b&z&0\\
	0&{{z^ * }}&c&0\\
	{{w^ * }}&0&0&d
	\end{array}} \right).
\end{equation}
These states  are characterized by seven real parameters.
In a complex X-state (\ref{x shape}),
without changing other parameters and affecting quantum correlations,
one can make the parameters $w$ and $z$ real by a local unitary transformation, $\rho_X\rightarrow
(U_{A}\otimes U_B)\rho_X(U_A\otimes U_B)^\dagger$ where $U_A$ and
$U_B$ are diagonal unitaries.
Henceforth, we can restrict ourselves to simulate real X-states as long as quantum correlation  is the subject of investigation.\\

\noindent X-states frequently appear in different contexts of physics 
\cite{YE07,WCRG10,S09,CRC10,KAM08,MWFC10}. They are of special importance due to their robustness in almost every noisy environment \cite{YE07},  and since a general two-qubit
state is mapped into this set under a noisy channel \cite{MWFC10}.
Moreover, the set of X-states can recover the entire spectrum and 
entanglement which are available for a general two-qubit state.
Indeed, two-qubit X-states enjoy entanglement universality, i.e.
an arbitrary two-qubit state can be mapped to its X-state counterpart 
by applying an entanglement preserving unitary transformation \cite{H18,H13,MMG14,MHGM15,MMH17}.\\

Here, to further physically motivate this study, we consider a physical model which is naturally relevant to the X-states. Such a model is the Heisenberg XYZ two-spin system in an inhomogeneous magnetic field where by changing its parameters, i.e. the magnetic field, its inhomogeneity and the strength of the spin couplings, a fairly large subset of X-states can be covered.  This gives us the opportunity to simulate the time evolution of a physically generated
X-state for various physical parameters, i.e. coupling constants.\\

\noindent The usual procedure for simulating  the dynamics of a physical model governed by a Hamiltonian $H$ consists of simulating three different modules, where from left to right, the first prepares the initial state, the second simulates the dynamics and the last module simulates the measurements of interest. It is also possible to merge these different parts to simplify its implementation. 
The part which simulates the dynamics, through decomposition of $U=e^{-iHt}$ into simple one qubit and CNOT gates, is the one which uses the largest number of gates. In particular, if the Hamiltonian is a sum of non-commuting terms, one has to use a Suzuki-Trotter approximation which drastically increases the number of gates. This is certainly the ultimate goal of digital quantum simulation of many-body systems, where analytical results are difficult or impossible to obtain. However, here we are concerned with a model whose 
analytical results are available and it can be simulated on the 
IBM publicly-available-few-qubit quantum computer.
So we can also compare the exact and simulation results 
to assess the effect of noise and other imperfections on our algorithm.\\


\noindent It is in view of this comparison that we follow a very simple procedure which partly uses the analytical results in the simulation. In return we show how this model can be simulated by using  only two or four qubits with very few gates, where impressive agreement between the exact and simulation results can be obtained. We stress that the aim of this paper and indeed many of the problems which have been solved by the publicly available qubits in the IBM quantum computer, is not to simulate a problem which is otherwise intractable analytically, but to make a comparison between exact and simulation results. 
Of Course, there are other   methods such as randomized benchmarking \cite{MGE12}, quantum volume \cite{C19,javadi}, etc., by which one can 
determine the performance of a quantum computer. However, 
there are also other approaches which combine analytical and simulation results to explore the range of problems that can be solved on a quantum computer. Our method which is in the direction of \cite{G19,PM19} falls in this direction.  \\

\noindent The structure of this paper is as follows: In section (\ref{pre}) we show how the X-states can be simulated and show that their X-ness is robust against noise in the qubits and gates.  In section (\ref{sec:Heis}) we review the Heisenberg two spin model and obtain the exact results, in section (\ref{Heis.sim}), we simulate the model in the absence of external magnetic field, and measure the entanglement and its dependence on the parameters of the initial state and the couplings of the Hamiltonian. Finally in section (\ref{magnet}) we extend our results to the case where an inhomogeneous magnetic field is also present. This time we will see how entanglement depends on both the anisotropy of the couplings and the inhomogeneity of the magnetic field. We end the paper by a discussion.

\section{Simulation  of X-states}\label{pre}
\noindent
A real X-state can  be written as a classical mixture
\be\label{X}
\rho_X=\sum_{i,j=0}^1p_{i,j}|\psi_{ij}\ra\la \psi_{ij}|,
\ee
of partially entangled states
\begin{align}\label{xbasis}
&\ket{\psi_{00}}=\cos\theta\ket{00}+\sin\theta\ket{11},
\quad\quad\quad\nonumber\
\ket{\psi_{01}}=\sin\phi\ket{01}+\cos\phi\ket{10},&\\
&\ket{\psi_{10}}=\cos\phi\ket{01}-\sin\phi\ket{10},
\quad\quad\quad
\ket{\psi_{11}}=-\sin\theta\ket{00}+\cos\theta\ket{11}.&
\end{align}
\begin{figure}[t]
	\centering\vspace{-1cm}
	\includegraphics[scale=0.3]{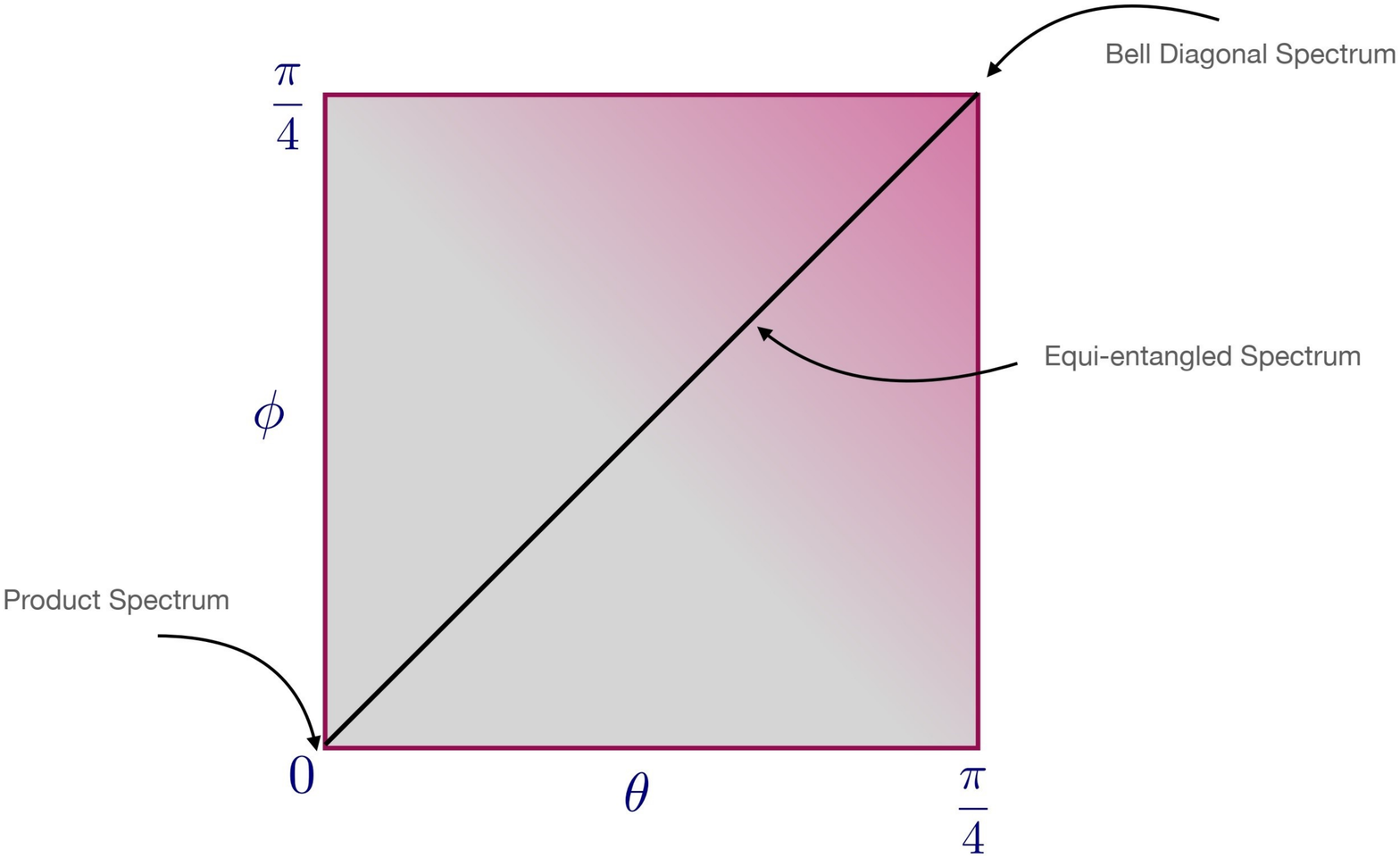}\vspace{-1cm}
	\caption{Characterizing the X states according to their spectrum.
		The Bell diagonal states correspond to a single point.
		The diagonal corresponds to equi-entangled spectrum.
		The four sides correspond to the states for which only two of the
		eigenstates are either product states or maximally entangled ones. }
	\label{e1}
\end{figure}
Therefore, the five parameters of a real X state are embodied in the three parameters of the classical mixture and the angles $\theta$ and $\phi$ which characterize the entanglement of the above states. The manifold of all real X stats is then the direct product of the 3-simplex (tetrahedron)   of probabilities and the square shown in figure \ref{e1}. The Bell diagonal states of Eq. \eqref{bds}, studied in \cite{G19}, correspond to a single point of this square. The anti-diagonal line of the square corresponds to a mixture of equi-entangled states \cite{KM06}, where all the states in (\ref{xbasis}) have the same amount of entanglement and interpolate between a mixture of product states and the Bell diagonal states.
\\

\begin{figure}[h]
	\centering\vspace{-.5cm}
	\includegraphics[scale=0.6]{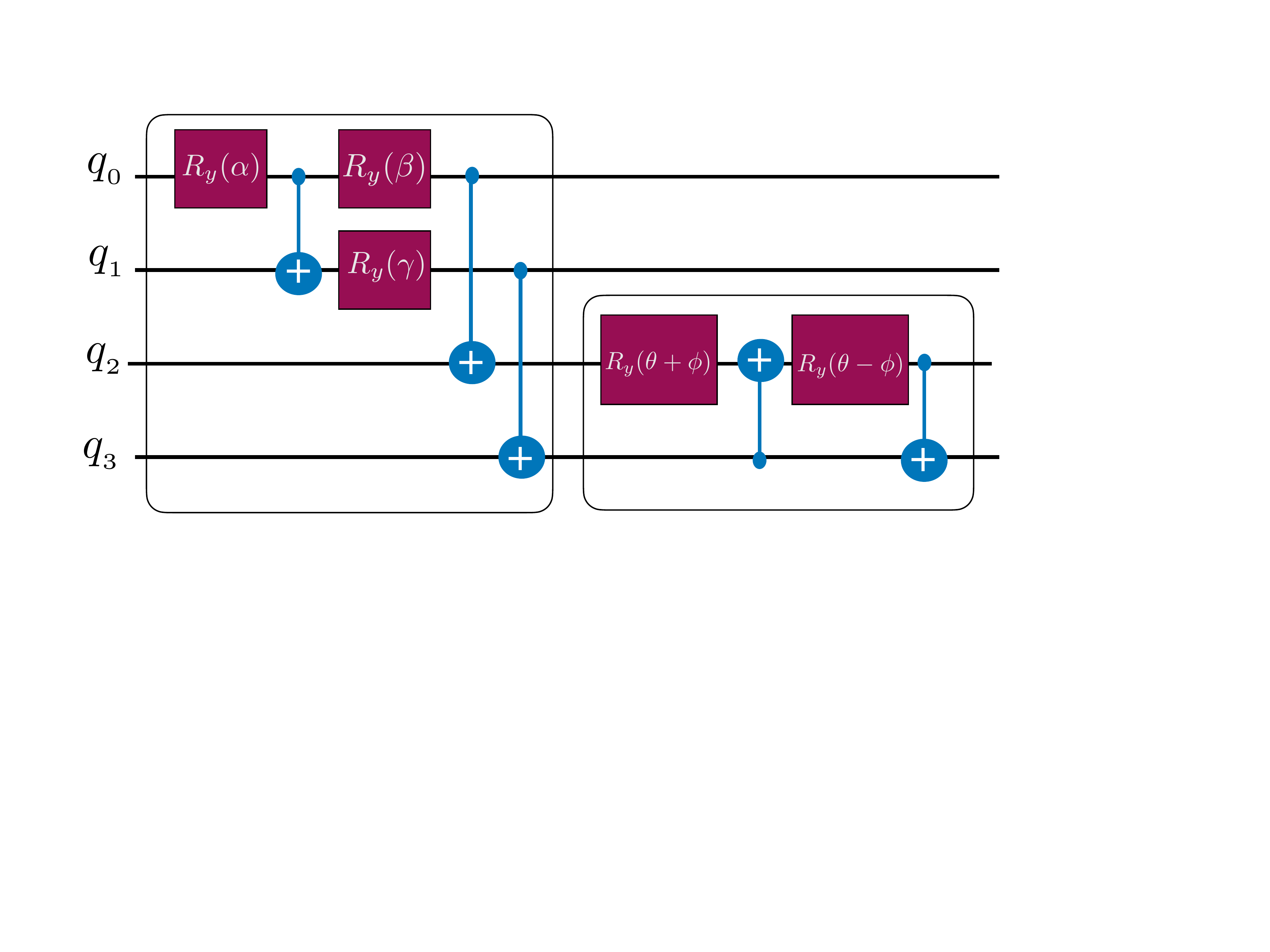}{\vspace{-4cm}}
	\caption{Quantum circuit for generating a general real X-state.
		The first block is to induce a classical state
		to the last two qubits. The second block
		sends this classical state into a real X-state.
		To generate a general complex X-state, one needs finally 
		apply two local and diagonal $R_z$ gates, one on each of the
		third and fourth qubits. 
	}
	\label{c1}
\end{figure}

\noindent These real X-states are generated by the circuit of figure (\ref{c1}).
 The first block in this circuit,  with appropriate tuning of the three angles $(\a,\beta, \gamma)$ \cite{G19}, produces a pure state
$$|P\ra=\sum_{i,j}\sqrt{p_{i,j}}|i,j\ra|i,j\ra$$ on the four qubits which leads to a mixed state
\be\label{classic}
\rho=\sum_{i,j}p_{i,j}|i,j\ra\la i,j|
\ee
on the third and fourth qubits.
 The second block now evolves the product basis $\{\ket{i,j}\}$ to the basis
$\{\ket{\psi_{i,j}}\}$ in Eq. \eqref{xbasis}, such that
\begin{equation}\label{ux1}
U(\theta,\phi)\ket{i,j}=\ket{\psi_{i,i\oplus j}},
\end{equation}
where $U(\theta,\phi)$ denotes the two-qubit  unitary operator of the second block acting on the third and fourth qubits. The states $\ket{\psi_{i,j}}$ are in fact the eigenstates of a real X-state, where $|\psi_{00}\ra$ and $|\psi_{11}\ra$ pertain to the outer block (or the even parity sector) and $|\psi_{01}\ra$ and $|\psi_{10}\ra$ pertain to the inner block (the odd parity sector). The whole circuit produces a real X-state with parameters
\begin{eqnarray}\label{spec-shape}
&a=p_{00}\cos^2\theta+p_{10}\sin^2\theta,
\quad\quad\quad\nonumber
d=p_{00}\sin^2\theta+p_{10}\cos^2\theta,&\\
&b=p_{01}\sin^2\phi+p_{11}\cos^2\phi,
\quad\quad\quad\nonumber
c=p_{01}\cos^2\phi+p_{11}\sin^2\phi,&\\
&w=(p_{00}-p_{10})\cos\theta\sin\theta,
\quad\quad\quad
z=(p_{01}-p_{11})\cos\phi\sin\phi,&
\end{eqnarray}

defined in Eq. \eqref{x shape}.
Appendix \ref{parameters} explains how to solve these equations for a given X-state.
 If  we set $\theta=\phi$, pertaining to the opposite diagonal of the square in Fig. \ref{e1}, we get a subclass of X-states
characterized by equi-entangled basis \cite{KM06}.  In this case the states $\ket{\psi_{ij}}$ all have the same value of
entanglement and interpolate between a product basis for $\theta=\phi=0$
to a maximally entangled basis for $\theta=\phi=\frac{\pi}{4}$.
When we further restrict the parameters to  $\theta=\phi=\frac{\pi}{4}$, i.e. a point in the square of Fig. \ref{e1}, the resulting X-state is a mixture of Bell states and is called a Bell diagonal state. These types of states have already been simulated on the IBM quantum computer
\cite{PM19,G19}. Moreover, to get a general complex X-state, the circuit of Fig. \ref{c1} should be accompanied by two local $R_z$ gates at the end, each on one of the third and fourth qubits.\\

It is worth mentioning that the circuit introduced in Fig. \ref{c1} 
is not the only possible set of gates one can apply 
to generate X-states.
Indeed, there are many other realizations as well.
For example,  the second block in the above circuit can be replaced
by the following set of gates
\begin{equation}\label{ux2}
U'(\theta,\phi)=C_1C_2\left(R_y(\theta-\phi)\otimes I\right)C_2
\left(R_y(\theta+\phi)\otimes I\right),
\end{equation}
 where the two-qubit operators $C_1$ and $C_2$ are CNOT gates
 whose control qubits are respectively the first and the second qubits.
$U'(\theta,\phi)$ then generates the same X-state as $U(\theta,\phi)$ of
Eq.  \eqref{ux1} with a change in parameter $\phi\rightarrow\frac\pi2-\phi$
which does not affect the generality of the simulated X-states.
Concerning the current difficulties in applying joint quantum gates
the gate decomposition of \eref{ux2} does not offer a more profitable
setup  with respect to Fig. \ref{c1} at the consequence of an extra CNOT.
However, once the simulation of a mixture of equi-entangled basis is
of interest, such a decomposition turns out to be more efficient.
Note that in this case we should set $\theta=\phi$, so that
$R_y(\theta-\phi)=R_y(0)=I$. Substituting this in Eq. \eqref{ux2}, we find
two CNOT gates $C_2$ after and before $R_y(\theta-\phi)$ cancel each other
leaving \eref{ux2} with only one CNOT gate $C_1$.
This implies while for simulating of a general X-state Fig.~\ref{c1} suggests
a simpler gate design, to prepare the measure zero subset of states
diagonal in the equi-entangled basis, \eref{ux2} is more efficient.
Applying the gate decomposition of \eqref{ux2}, if we set
$\theta=\phi=\frac\pi4$ the circuit generating BDSs \cite{G19} is recovered.
We will also introduce in the sequel other possibilities for simulating an X-state.\\

\begin{figure}[h]
	\hspace*{-1cm}
	\includegraphics[scale=0.48]{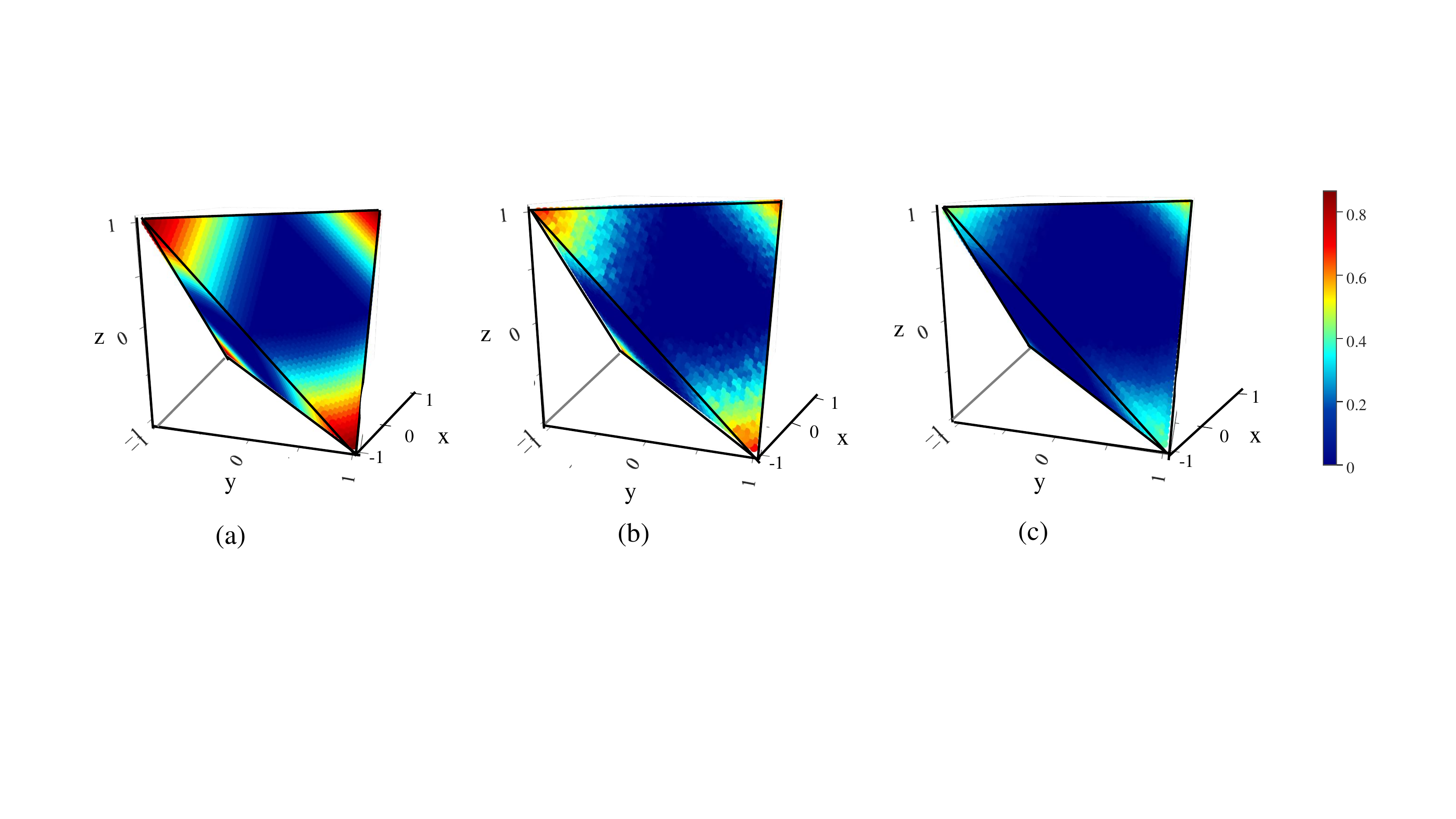}\vspace{-3cm}
\centering
	\caption{(a) Analytical, (b) Noisy simulation via Qiskit, and (c) IBM results for concurrence of a mixture of equi-entangled basis with $\theta=\phi=\frac\pi6$, see \eref{xbasis}. Simulation results are obtained from IBM-Valencia device on $2020-12-25$ with $8000$ shots}
	\label{t1}
\end{figure}

\noindent A comparison between theoretical results (i.e. entanglement, discord, steering) on the one hand and simulation results obtained by Qiskit and the
real IBM device on the other hand has already  been made for the BDSs \cite{G19}. It has been shown that a pretty good agreement between these three kinds of results exists as long as we are not near the edges or corners of the tetrahedron of probabilities.  The authors of \cite{G19} show that as one moves toward the edges of the tetrahedron  of BDSs, the noise model of Qiskit becomes more and more insufficient to mimic the noise in the real quantum device and for some states  the drop in fidelity can be come as large as 70 percent.  Therefore to make a sensible comparison for X-states,
we do the same calculations concerning the concurrence for an X-state here.
We remind the readers  for a general X-states in Eq.~\eqref{x shape}
the concurrence  is given by
\begin{eqnarray} \label{con:x}
C(\rho )=2\max\{0,|w|-\sqrt{bc},|z|-\sqrt{ad}\}.
\end{eqnarray}
For simulation we restrict ourselves to a mixture of equi-entangled basis
specified by $\theta=\phi=\frac\pi6$ \cite{code}. Such a set of states, corresponding to a single point on the diagonal of the square \ref{e1}, can be visualized by the $3$-simplex, the tetrahedron of four dimensional probabilities appearing in the convex combination of Eq. \eqref{X}.
The results are shown in Fig. \ref{t1} in which  the analytical
amount of concurrence can be compared with the results obtained with noisy simulation on Qiskit and simulation on the actual hardware of IBM.
We see the farther from the vertices and edges the states are, the less deviation in concurrence can be obtained.\\

Note that in this set of equi-entangled states, 
like Bell diagonal ones, the states generally become more and more entangled as we go from the center of tetrahedron 
toward the edges and vertices.  
A random noise is not usually supposed to generate entanglement but 
to destroy it, which explains why there is more agreement
in the center of the tetrahedron where separable states are present. It is worth mentioning there are 
several methods to mitigate noises in a quantum computer
\cite{MZO20,BSKMG21,FHJKSW20,GPD20,CXB20,WE16,U21}.

\subsection{Robustness of X-states under noise}\label{robustness}
\noindent It has been claimed in \cite{YE07} that the X-states are robust under most kinds of noise in the environment. It is thus desirable to check this claim in simulation of these states in the IBM quantum computer. More precisely, we want to see if the circuit shown in Fig. \ref{c1} which is supposed to produce an X-state at the output, when run on the IBM quantum computer actually produces an X-state or not. In principle, one should check to see if all the non-X entries of $\rho$, like $\rho_{00,01}$ are zero.
This can be examined once the tomography process is applied
on the output state. The usual approach to do state tomography is to
reconstruct the state by obtaining the expectation values of the form
$\Tr[\rho(\sigma_i\otimes\sigma_j)]$ for $i,j\in\{0,\dots,3\}$,
where $\sigma_0=I$ and others are Pauli matrices.
This gives us $15$ independent parameters required to specify a state.
In the lab, however, one has access to probability distributions
of a measurement rather than the expectation values.
This helps to get the marginal probability distributions from the joint ones
instead of measuring the local observables separately.
So the full state tomography can be done by measuring nine observables
$\{\sigma_i\otimes\sigma_j\}$ where $i,j\in\{1,2,3\}$, which is the
approach that is already adopted by IBM.
In case the output state remains an X-state, one can reconstruct the state
by applying five measurements rather than nine which are
$\{\sigma_1\otimes\sigma_1,\sigma_2\otimes\sigma_2,
\sigma_3\otimes\sigma_3,\sigma_1\otimes\sigma_2,
\sigma_2\otimes\sigma_1\}$.
Moreover, if we can make sure that the output states are close enough to a
real X-state, then the state reconstruction can be done by a
tomography process including measuring three observables
$\{\sigma_i\otimes\sigma_i\}$ for $i\in\{1,2,3\}$.
\\

The  fact that full tomography process may be replaced by a partial one
results in a significant reduction in run-time of the algorithms.
Furthermore, to apply different measurements, one has to add more gates
which in turn increases the noise and the time evolution of the system.
So the partial tomography is expected to also reduce the noise on the systems.
This is  however possible if the (real) X-states are robust under the
environmental noise and gate imperfections, i.e. a general (real) X-state
remains a (real) X-state. To examine this fact hereafter we will apply
a full tomography process, a partial tomography process with five
measurements, and a partial tomography process with three measurements
when we expect  real X-states as the output states, see figures \ref{fid1} and \ref{fid2}.\\

Let us denote by $F^f$ the fidelity between analytical results and those
obtained by applying full tomography to the output states.
Similarly $F^p_5$ and $F^p_3$ are to show the  fidelity between analytical
results and the states reconstructed by a tomography process with $5$ and $3$
musearments, respectively.
Our results  suggest the maximum difference between $F^p_5$ and
$F^p_3$ is $0.05$, while in $70\%$ of the studied cases this difference
is less than $0.01$. This implies  a real X-state, with an acceptable precision,
is as robust as an X-state itself.
On the other hand, the maximum difference of $F^p_5$ and
$F^f$ is $0.09$, while in $60\%$ of cases this difference
is less than $0.01$ and for $80\%$ of cases this difference
is less than $0.05$.
This difference might be a consequence of partial tomography and reduction
in noise due to decreasing gates and time evolutions as mentioned above.
These results indicate that (real) X-states are approximately
robust under noise of the IBM hardware.
\\

\noindent We now turn to a physical model whose Hamiltonian and evolution operator are in X-form and hence produces X-states for a large class of initial states. The role of the probabilities is now played by the parameters of the initial state and the roles of $\theta$ and $\phi$ are played by the coupling constants of the Hamiltonian and time. We first consider the Heisenberg XYZ two-spin system and later put this system in an inhomogeneous magnetic field.

\section{The Heisenberg XYZ spin system: Analytical results}\label{sec:Heis}

\noindent Consider two spin one-half particles subjected to the Hamiltonian
\begin{equation}\label{xyz}
H = \frac{1}{2}\left( {J_x}\sigma_x\otimes\sigma _x+
{J_y}\sigma _y\otimes\sigma _y + {J_z}\sigma _z\otimes\sigma _z \right)
\end{equation}
where ${J_\mu }\left( {\mu  = x,y,z} \right)$ are the real coupling constants,
with positive values for antiferromagnetic phase and
negative values for  ferromagnetic case. This Hamiltonian has the symmetry
\be
[H, \sigma_z\otimes \sigma_z]=0,
\ee
which is in fact the defining relation of a matrix to be in X-shape:
\begin{equation}\label{hamiltonian}
H =  \left( {\begin{array}{*{20}{c}}
	{\frac{1}{2}{J_z} }&0&0&{J\kappa }\\
	0&{ - \frac{1}{2}{J_z} }&J &0\\
	0&J &{ - \frac{1}{2}{J_z} }&0\\
	{J\kappa }&0&0&{\frac{1}{2}{J_z} }
	\end{array}} \right).
\end{equation}
Here $J= \frac{{{J_x} + {J_y}}}{2}$,
$\kappa  = \frac{{{J_x} - {J_y}}}{{{J_x} + {J_y}}}$. Therefore, the parameter $\kappa$ characterizes the amount of anisotropy of the couplings in the $x-y$ plane. \\

\noindent The X-shape of the Hamiltonian or its symmetry means that the two subspaces of even and odd parity,
spanned respectively by $\{|00\ra,|11\ra\}$ and $\{|01\ra, |10\ra\}$, evolve independently. This symmetry remains intact when we later add an inhomogeneous magnetic field in the z-direction.
The energy eigenvalues are given by
\be\label{eigenvalue}
{\varepsilon _{1,2}} = \frac{1}{2}{J_z} \pm J\kappa, \h
{\varepsilon _{3,4}} = -\frac{{ 1}}{2}{J_z} \pm J,
\ee
corresponding to the following eigenvectors,
respectively
\begin{eqnarray}\label{eigenvector}
\ket{\Phi_{1,2}}= \frac{1}{\sqrt{2}}\left( \ket{00}
\pm\ket{11}\right),\hspace{1cm}
\ket{\Phi_{3,4}}\frac{1}{\sqrt{2}}\left( \ket{01}\pm
\ket{10}\right).
\end{eqnarray}
Straightforward calculations show that the evolution operator $U=e^{-iHt}$ is given by
\begin{equation}\label{evolution1}
U(t)  = e^{-\frac{i}{2}J_zt}
\left( {\begin{array}{cc}
	\cos J\kappa t &  -i\sin J\kappa t\\  -i\sin J\kappa t&\cos J\kappa t
	\end{array}} \right)_{1,4}\oplus e^{\frac{i}{2}J_zt}
\left( {\begin{array}{cc}
	\cos Jt & -i\sin J t\\ -i\sin J t&\cos Jt
	\end{array}} \right)_{2,3}
\end{equation}
in which the subscripts indicate the position of the block, note that the (1,4) block corresponds to the even sector and the (2,3) block to the odd sector.
The symmetry and the resulting invariance, allow us to consider  physically important class of initial states and consider their dynamics separately. This restriction is worth the immense simplification in the IBM simulation as we will see.  \\

\noindent Consider an initial classically correlated state in the even parity sector ($\{|00\ra,|11\ra\}$)
\be \label{initial}
\rho(0)=\frac{1+\lambda}{2}|00\ra\la 00|+\frac{1-\lambda}{2}|11\ra\la 11|,
\ee where $-1\leq \lambda \leq 1$.
 This state evolves  to
 \be\label{f1}
 \rho(t)=\frac{1}{2}\left(\begin{array}{cc} 1+\lambda \cos 2J\kappa t & i\lambda \sin 2J\kappa t \\ -i\lambda \sin 2J\kappa t & 1-\lambda \cos 2J\kappa t  \end{array}\right)_{14},
 \ee
 where by the subscript $(1,4)$ we mean that this density matrix should be embedded into the outer block of the full two-qubit density matrix.
Therefore we find the concurrence of this state \eqref{con:x} to be
\be\label{ct}
C(t)=|\lambda||\sin 2J\kappa t| .
\ee
The interesting point is the factoring of this quantity into $|\lambda|$, which pertains only to the initial state, and $|\sin 2J\kappa t|$,  which comes solely from the dynamics of the XYZ spin pair. But what is the significance of $|\lambda|$? It is clear from (\ref{initial}) that the original state $\rho(0)$ has no entanglement, no discord, neither any coherence in the computational basis. However when written in the Bell basis with $\ket{\phi_\pm}=\frac{1}{\sqrt2}(\ket{00}\pm\ket{11})$, we find
\be
\rho(0)=\frac{1}{2}(|\phi_+\ra\la \phi_+|+|\phi_-\ra\la \phi_-|)+\frac{\lambda}{2}(|\phi_+\ra\la \phi_-|+|\phi_-\ra\la \phi_+|).
\ee
This means that $|\lambda|$ is the coherence of the initial state in the Bell basis measured by the $l_1$-norm of coherence \cite{BCP14}. Therefore, what equation (\ref{ct}) tells us is that the dynamics of XYZ chain, evolves the initial coherence of the state into entanglement at later times. As far as the dynamics is concerned, we see that  the amount of entanglement is controlled by the degree of anisotropy $J\kappa$ and changes in a periodic fashion. In fact  in the absence of anisotropy, i.e. when $\kappa=0$, no entanglement can be generated over time. \\

\noindent The same considerations are true when the initial state is in the odd parity sector and is of the form
$\rho(0)=\frac{1+\mu}{2}|01\ra\la 01|+\frac{1-\mu}{2}|10\ra\la 10|$ which is now evolved to
\be\label{f2}
\rho(t)=\frac{1}{2}\left(\begin{array}{cc} 1+\mu \cos 2J t & i\mu \sin 2J t \\ -i\mu \sin 2J t & 1-\mu \cos 2J t  \end{array}\right)_{23}.
\ee
Proceeding as before, one now finds that
\be\label{cf2}
C(t)=|\mu||\sin 2J t|.
\ee
Again  the entanglement is factorized into two parts, a part which comes from initial coherence $|\mu|$ in the $\{\ket{\psi_+},\ket{\psi_-}\}$ subspace with $\ket{\psi_\pm}=\frac{1}{\sqrt2}(\ket{01}\pm\ket{10})$, and a part which comes from the dynamics of XYZ spin system. This time, however, the anisotropy does not play a role and we can have entanglement even for isotropic couplings.  \\

\noindent Before adding a magnetic field, it is instructive to see how this physical system can be simulated on the IBM quantum computer. We do this in the next section and later on we will show how this simulation  should be modified in order to incorporate the magnetic field.

\section{The Heisenberg XYZ spin system: Simulation results}\label{Heis.sim}
\noindent We now show that with two qubits of the IBM quantum computer, we can simulate the Heisenberg system and measure  entanglement of the dynamical density matrices \eqref{f1} and \eqref{f2}. We use the invariance of the even and odd parity sectors and study them separately.
 The point is that the entanglement of the state (\ref{f1}) is the same if we replace the evolution operator in (\ref{evolution1})  with a real one, namely $R_y(\theta)=\left(\begin{array}{cc}\cos\frac\theta2 &-\sin\frac\theta2 \\ \sin\frac\theta2 & \cos\frac\theta2 \end{array}\right)=e^{i\frac\theta2 \sigma_y}$. This will lead to a real density matrix with the same amount of entanglement. This modification will show its full simplification when we consider both sectors in section (\ref{magnet}).\\
\begin{figure}[H]
	\vspace*{-1cm}
	\includegraphics[scale=.35]{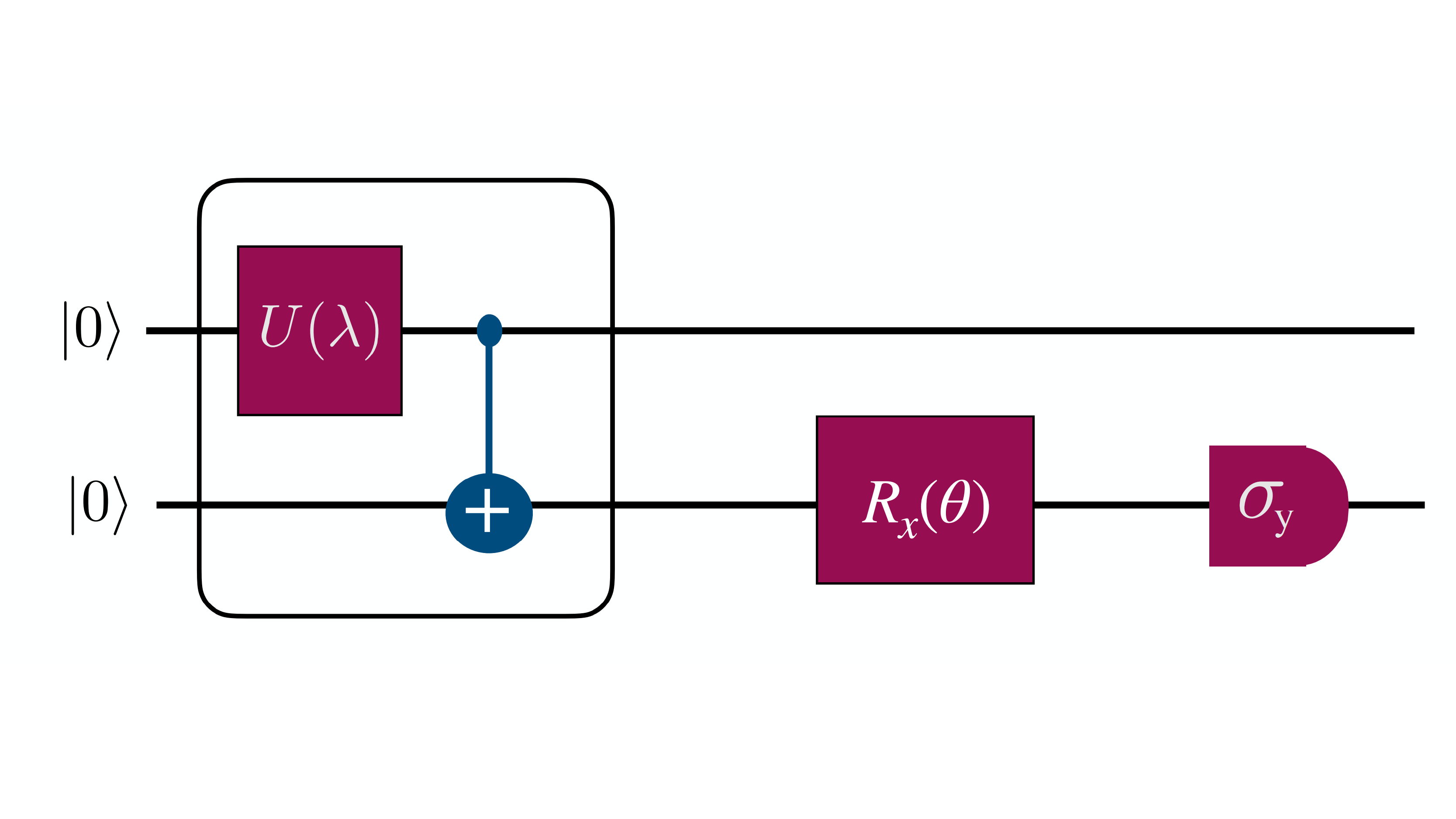}\vspace{-2cm}
\centering
	\caption{The first module prepare the mixed state (\ref{initial}) on the second qubit. The rest evolves this state according to the XYZ Hamiltonian. The expectation value of the measurement at the end is proportional to the concurrence \eqref{ct}.  }
	\label{c1newnew}
\end{figure}
\noindent Consider the circuit shown in figure (\ref{c1newnew}), where
\be
U(\lambda)=\frac{1}{\sqrt{2}}\left(\begin{array}{cc}\sqrt{1+\lambda}&-\sqrt{1-\lambda}\\ \sqrt{1-\lambda}&\sqrt{1+\lambda}\end{array}\right)\h
R_x(\theta )=\left(\begin{array}{cc}\cos \frac\theta2 &-i\sin\frac\theta2 \\ -i\sin\frac\theta2 &\cos\frac\theta2 \end{array}\right).
\ee
These gates are nothing but rotations around the $y$ and the $x$ axis and are among the allowable gates in Qiskit and the IBM quantum computer. The first part of this circuit produces the state
\be
|\psi\ra_{1}=\sqrt{\frac{1+\lambda}{2}}|00\ra+\sqrt{\frac{1-\lambda}{2}}|11\ra
\ee
on the first two qubits. The unitary operator $R_x(\theta)$, produces a state which, after ignoring the first qubit,  leaves us with the following mixed state on the second qubit
\be
\rho=\frac{1+\lambda}{2}|\phi_{0}\ra\la \phi_{0}|+\frac{1-\lambda}{2}|\phi_{1}\ra\la \phi_{1}|,
\ee
where
\be
|\phi_0\ra=\cos\frac\theta2 |0\ra+i\sin\frac\theta2 |1\ra\h |\phi_1\ra=-i\sin \frac\theta2 |0\ra+\cos\frac\theta2 |1\ra
\ee
The last part, measures $|Tr(\sigma_y\rho)|=|\lambda \sin\theta t|$, which when $\rho$ is embedded as in the (1,4) block is nothing but the concurrence of the two-qubit output state (\ref{f1}). Of course we have to set $\theta=2\kappa J t$ in this sector.  \\

\noindent One can now change $\theta$ from $2J\kappa t$ to $2J t$ and run the same circuit as before to find the concurrence of the state (\ref{f2}), when only the inner block is non-zero. This concurrence is given by (\ref{cf2}). \\

\noindent If we want to simulate the dynamics, when both the odd and the even parity sectors are involved, we need to add two more qubits and use the circuit shown in figure (\ref{c1}). While in principle it is possible to prepare any classically correlated state like $\rho(0)=\sum_{i,j}p_{ij}|i,j\ra\la i,j|$  with three parameters $p_{ij}$  \cite{G19}, for simplicity, we take the initial state to be of the type
\be\label{initial2}
\rho(0)=\left(\frac{1+\lambda}{2}|0\ra\la 0|+\frac{1-\lambda}{2}|1\ra\la 1|\right)\otimes \left({\cos ^2}(\frac{\gamma }{2}) |0\ra\la 0|+ {\sin ^2}(\frac{\gamma }{2})|1\ra\la 1|\right).
\ee


 Consider the second block in figure (\ref{c1}), where $\theta=J\kappa t$ and $\phi=Jt$. As mentioned, this block affects the following transformation (note the mismatch of subscripts),
$\ket{i,j}\lo\ket{\psi_{i,i\oplus j}}$ where $\ket{\psi_{i,j}}$'s are given by \eref{xbasis}. As it is seen this is not the way that these states evolve under the dynamics of the Heisenberg chain, i.e. it will be so if we implement a CNOT gate before the middle block. \\

To apply circuit \ref{c1} for simulating the two-qubit Heisenberg evolution of the initial state \eqref{initial2} we should set in the first block
\begin{equation}
\alpha=0,\quad\beta=2\arccos\sqrt\frac{1+\lambda}{2}.
\end{equation}
Note that taking $\alpha=0$ neutralizes the first rotation gate. As a consequence we can also drop the first CNOT between the first two qubits since it does not affect the input of the circuit, i.e. $C_1\ket{00}=\ket{00}$. This, however, holds once the initial state of the evolution is a product state as the one in \eref{initial2}.
\noindent The combination of the two blocks now produces the state
\ba\label{out1}
\rho(t)&=&{\cos ^2}(\frac{\gamma }{2})\left[\frac{1+\lambda}{2}|\psi_{00}\ra\la \psi_{00}|+\frac{1-\lambda}{2}|\psi_{11}\ra\la \psi_{11}|\right]\nonumber\\&+&{\sin ^2}(\frac{\gamma }{2}) \left[\frac{1+\lambda}{2}|\psi_{01}\ra\la \psi_{01}|+\frac{1-\lambda}{2}|\psi_{10}\ra\la \psi_{10}|\right]
\ea
which is a real density matrix with the same amount of entanglement as the one produced by the Heisenberg Hamiltonian. \\

\begin{figure}[H]
\hspace*{-0.75cm}
\includegraphics[width=14cm,height=8cm]{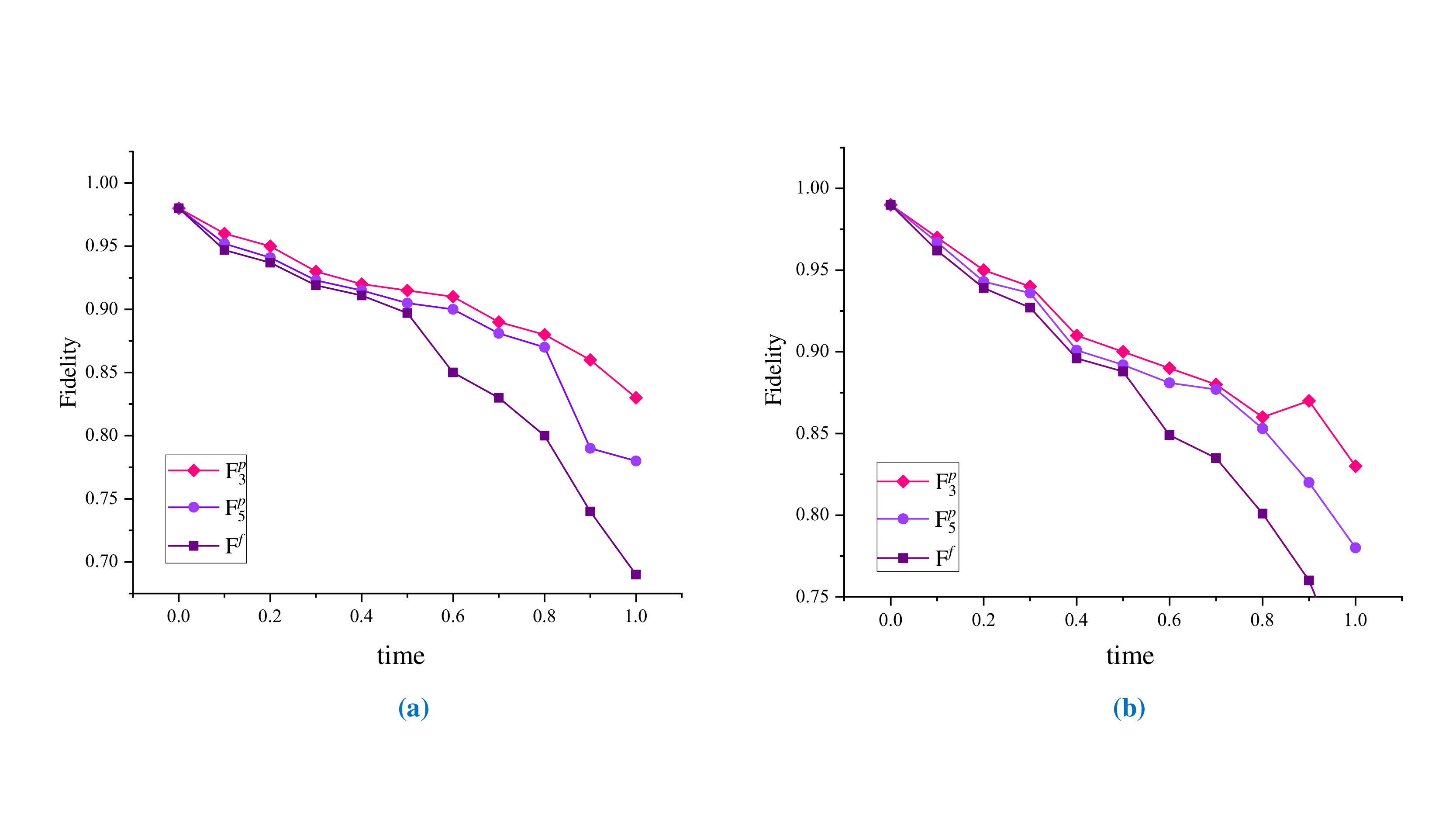}\vspace{-1cm}
\centering
\caption{(Color online) The part (a) shows the fidelity obtained between the state \eqref{out1}, where $\lambda=1$, $\gamma=2\arccos\sqrt{7/8}$, $J=1$, and $\kappa=0.75$, and the simulation results obtained via full tomography process $(F^f)$, partial tomography process by $5$ observables $(F^p_5)$, and partial tomography by $3$ observables $(F^p_3)$, see Section \ref{robustness}.
Panel (b) shows the same quantities for the state obtained by  $\lambda=0.9$, $\gamma=2\arccos\sqrt{3/4}$, $J=1$, and $\kappa=0.75$.
Simulation results are obtained from ibmq-manila device on $2021-05-23$ with $8000$ shots. }
\label{fid1}
\end{figure}

\begin{figure}[!h]
\hspace*{-0.75cm}
\includegraphics[scale=0.4]{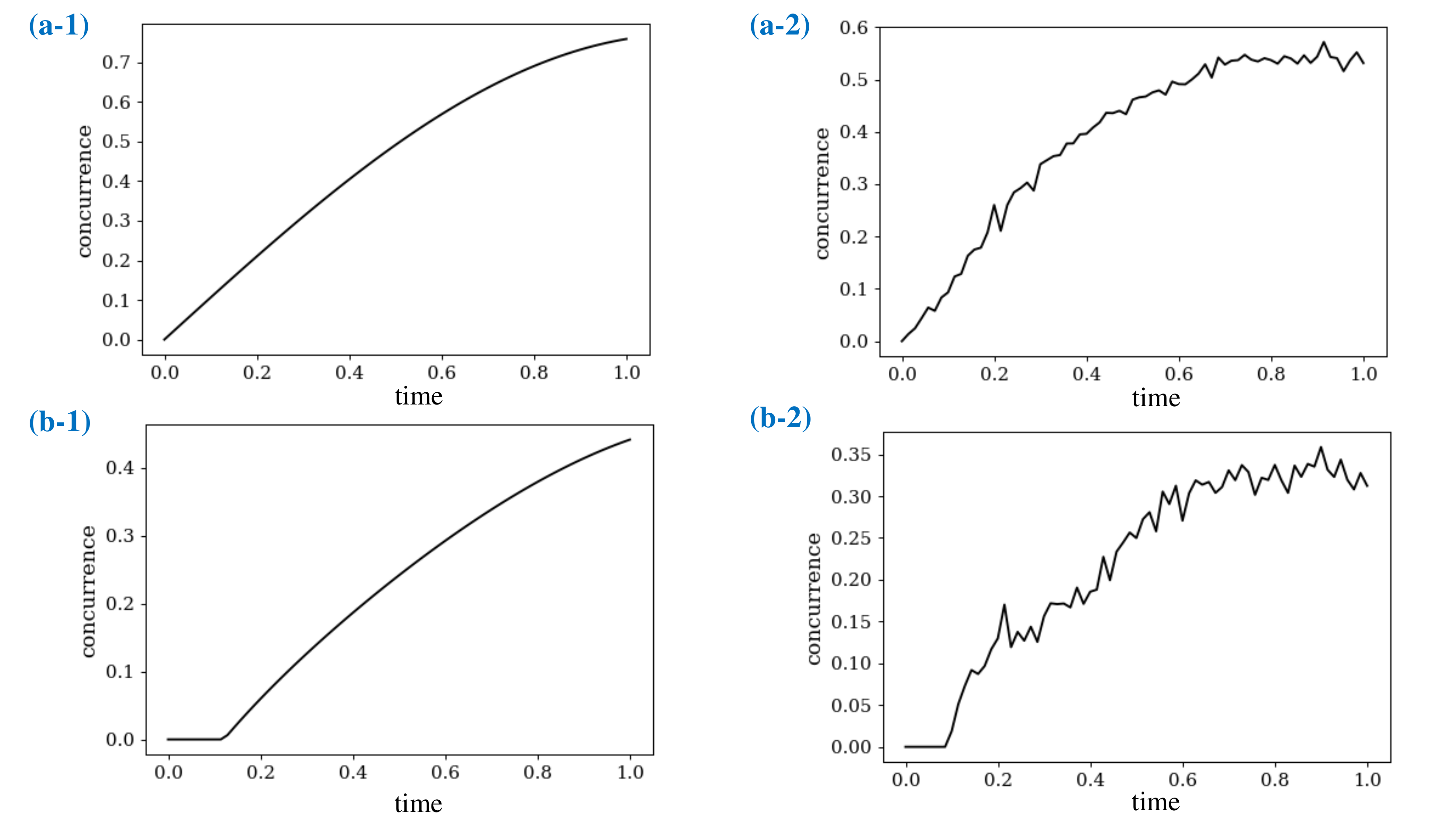}
\centering
\caption{The plots of panel (a) show the concurrence obtained for the state \eqref{out1}, where $\lambda=1$, $\gamma=2\arccos\sqrt{7/8}$, $J=1$, and $\kappa=0.75$, (a-1) analytically, and (a-2) by partial tomography process with $3$ observables, see Section \ref{robustness}.
Panel (b) shows the same quantities for the state obtained by  $\lambda=0.9$, $\gamma=2\arccos\sqrt{3/4}$, $J=1$, and $\kappa=0.75$.
Simulation results are obtained from ibmq-manila device on $2021-05-23$ with $8000$ shots. }
\label{conc}
\end{figure}

To analyse this model we simulate it on the actual IBM quantum device
for several values of $\g$ and $\lambda$ and fixed values of $J$ and $\kappa$. To this end we first checked the robustness of the output states by applying the notion of fidelity as discussed in Section \ref{robustness}, see Fig. \ref{fid1}. In these plots we can compare
$F^f$ the fidelity between analytical results and the states obtained by full tomography process, $F^p_5$  the fidelity between analytical results and those obtained by partial tomography process with $5$ observables to show how X-states are robust, and $F_3^p$  the fidelity between analytical results and the outputs obtained by partial tomography process with measuring three observables to show how real X-states are robust.
We also present the concurrence of these states as  a function of time in
Fig. \ref{conc}. The simulation results are obtained by partial tomography process with three observables $\{\sigma_i\otimes\sigma_i\}_{i=1}^3$ as discussed in Section \ref{robustness}. In these plots we can compare analytical results with the simulation ones from ibmq-manila device.
It is seen in these figures now that the less entanged the states are, the more robustness can be observed.\\

\noindent {\bf Remark:} We stress again that the purpose of our work, like many other similar works \cite{G19,PM19} is to test the performance of a quantum simulation  in the real IBM quantum computer and compare it with the noisy model of Qiskit and the exact results. Otherwise, we could have avoided this 4 qubit circuit and use the decoupling of the two sectors, as discussed in previous section, and by combining the results of measurements in those sectors to determine the entanglement of the output density matrix.


\section{Inclusion of magnetic field}\label{magnet}

\noindent Let us now put the two spins in an inhomogeneous magnetic field, where the Hamiltonian will be
\begin{equation}\label{xyz}
H = \frac{1}{2}\left( {J_x}\sigma_x\otimes\sigma _x+
{J_y}\sigma _y\otimes\sigma _y + {J_z}\sigma _z\otimes\sigma _z +
{{\bm B}_1}\cdot\bm{ \sigma }\otimes I +
I\otimes{{\bm B}_2}\cdot\bm{\sigma }  \right),
\end{equation}
where ${B_j}\left( {j = 1,2} \right)$ is the magnetic field on site $j$.
 We let $
{\bm B_j} = {B_j}\hat z$ where ${B_1} = B + b$ and ${B_2} = B - b$.
Therefore, $B$ is the average magnetic field and $b$ is its inhomogeneity. The symmetry $[\sigma_z\otimes \sigma_z , H]=0$ remains intact and the Hamiltonian is again in X-shape, implying that the two subspaces pertaining to the inner and outer blocks evolve independently:
\begin{equation}\label{hamiltonian}
H =  \left( {\begin{array}{*{20}{c}}
	{\frac{1}{2}{J_z} + B}&0&0&{J\kappa }\\
	0&{ - \frac{1}{2}{J_z} + b}&J &0\\
	0&J &{ - \frac{1}{2}{J_z} - b}&0\\
	{J\kappa }&0&0&{\frac{1}{2}{J_z} - B}
	\end{array}} \right).
\end{equation}
 The  energy eigenvalues are given by
\be\label{eigenvalue}
{\varepsilon _{1,2}} = \frac{1}{2}{J_z} \pm\xi,\h
{\varepsilon _{3,4}} = -\frac{{ 1}}{2}{J_z} \pm \zeta,
\ee
corresponding to the eigenvectors,
\begin{eqnarray}\label{eigenvector}
\ket{\Phi_{1,2}}\propto \left( ({B\pm\xi})\ket{00}
+J\kappa\ket{11}\right),\hspace{1cm}
\ket{\Phi_{3,4}}\propto\left( ({b\pm\zeta})\ket{01}
+J\ket{10}\right),
\end{eqnarray}
where  $\xi= \sqrt {{B^2} + {J^2}{\kappa ^2}}$ and
$\eta=\sqrt {{b^2}  + {J^2}}$.
Straightforward calculations show that
\begin{equation}\label{evol}
U(t)  = e^{-\frac{i}{2}J_zt}
\left( {\begin{array}{cc}
	u&c\\ -c^*&u^*
	\end{array}} \right)_{1,4}\oplus e^{\frac{i}{2}J_zt}
\left( {\begin{array}{cc}
	u'&c'\\ -c'^*&u'^*
	\end{array}} \right)_{2,3}
\end{equation}
where
\be
u=\cos \xi t-i\frac{B}{\xi}\sin\xi t\h c= -i\frac{J\kappa}{\xi}\sin\xi t
\ee
and
\be
u'=\cos \eta t-i\frac{b}{\eta}\sin\eta t\h
c'= -i\frac{J}{\eta}\sin\eta t.
\ee
An initial state $\rho(0)=p|00\ra\la 00|+(1-p)|11\ra\la 11|$ now evolves to
\be\label{out3}
\rho(t)=\left(\begin{array}{cc} p|u|^2+(1-p)|c|^2& (1-2p)uc\\ (1-2p)u^*c^*& (1-p)|u|^2+p|c|^2 \end{array}\right)_{14}
\ee
whose concurrence, according to (\ref{con:x}),  is given  by $C(t)=2|1-2p||uc|$ or
\be
C(t)=2|\lambda|\frac{J\kappa}{\xi}|\sin\xi t| \sqrt{\cos^2\xi t+\frac{B^2}{\xi^2}\sin^2\xi t},
\ee
where we have parameterized $p=\frac{1+\lambda}{2}$ as before. Again we see a factorization of this concurrence into the coherence of the initial state in the Bell basis, as given by $|\lambda|$ and a part which comes solely from the dynamics. A modification of the previous circuit will allow us to simulate this quantum system and measure the final concurrence. The second module in circuit (\ref{c1newnew}) which simulates the unitary evolution should now implement the gate
\be
R_x(\theta)\longrightarrow V(t)= (U(t))_{14}=
\left( {\begin{array}{cc}
		u&c\\ -c^*&u^*
\end{array}} \right)=\left(\begin{array}{cc} \cos \xi t-\frac{iB}{\xi}\sin\xi t&-i\frac{J\kappa}{\xi}\sin \xi t\\ -i\frac{J\kappa}{\xi}\sin \xi t&\cos \xi t+\frac{iB}{\xi}\sin\xi t\end{array}\right).
\ee

\noindent Defining $\delta$ so that $\cos \delta =\frac{B}{\xi}$ and $\sin\delta =\frac{J\kappa}{\xi}$ (Note that $\xi=\sqrt{B^2+J^2\kappa^2}$), we see that
\be
V(t)=\cos\xi t I-i\sin\xi t \left(\begin{array}{cc} \cos \delta&\sin\delta\\ -\sin\delta &-\cos\delta\end{array}\right )=e^{-i\xi t {\bf n}\cdot{\bm \sigma}}
\ee
where ${\bf n}=\sin\delta \hat{\bf x}+\cos\delta \hat{\bf z}$. It is now easy to factorize the gate $U(t)$ into
\be
V(t)=e^{i\frac{\delta}{2}\sigma_y}e^{-i\frac \xi t\sigma_z}e^{-i\frac{\delta}{2}\sigma_y}.\ee
An exactly similar treatment applies to the dynamics in the odd sector. It is enough to change the triple ($\xi, J, B$) to   ($\eta,  J\kappa, b$) in the previous circuit to simulate the dynamics of the odd sector.  \\

Similar to the case without magnetic field, if the goal is to 
simulate the evolution of an initial state in the form of Eq.~
\ref{initial2}, then a circuit with four qubits in Fig.~\ref{c1} 
should be applied. Note that as it is mentioned in Section \ref{pre}, 
to generate the most general complex X-state, this circuit needs to be followed by two local $R_z$ gates on the third and fourth qubits at the end.
Once we are interested in quantum correlations, however, we may drop
these two local unitaries, and work with a real state with the same amount of entanglement.\\

\begin{figure}[h]
 \hspace*{-0.45cm}
\includegraphics[scale=0.5]{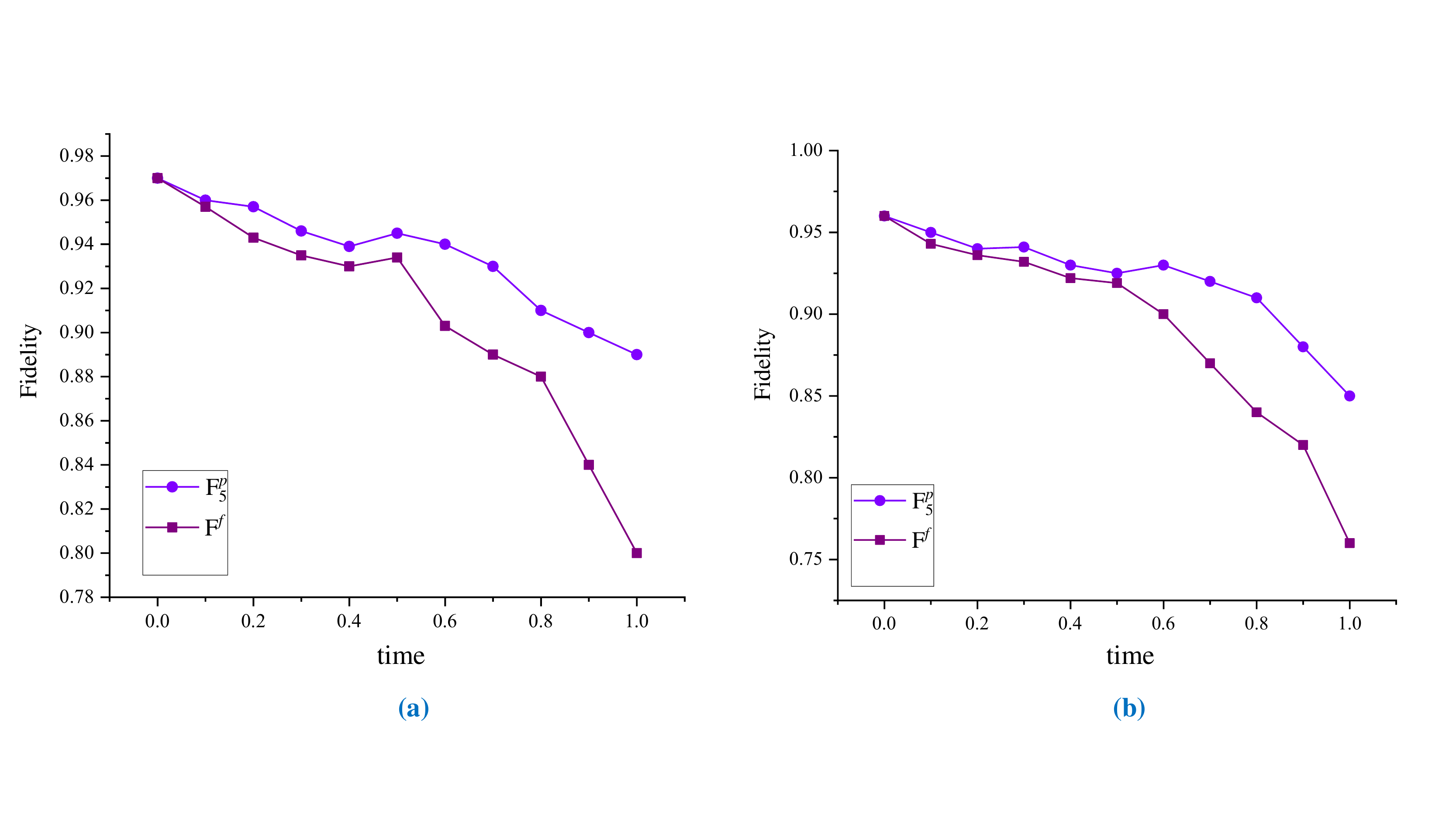}\vspace{-0.5cm}
\caption{(Color online) The fidelity between the state \eqref{initial2}  evolved under the evolution of \eqref{evol} and its simulated state reconstructed by full tomography process $(F^f)$, and partial tomography process by $5$ observables $(F^p_5)$, see Section \ref{robustness}. Here, $B=1$, $b=0.5$, $\lambda=1$, $\gamma=2\arccos\sqrt{3/4}$, $J=1$, and $\kappa=0.95$.
Panel (b) shows the same quantities for the state obtained once we change magnetic field and its inhomogenity as   $B=2.5$, and $b=0.25$.
Simulation results are obtained from ibmq-manila device on $2021-05-23$ with $8000$ shots. }
\label{fid2}
\end{figure}

\begin{figure}[!h]
\hspace*{-0.75cm}
\includegraphics[scale=0.43]{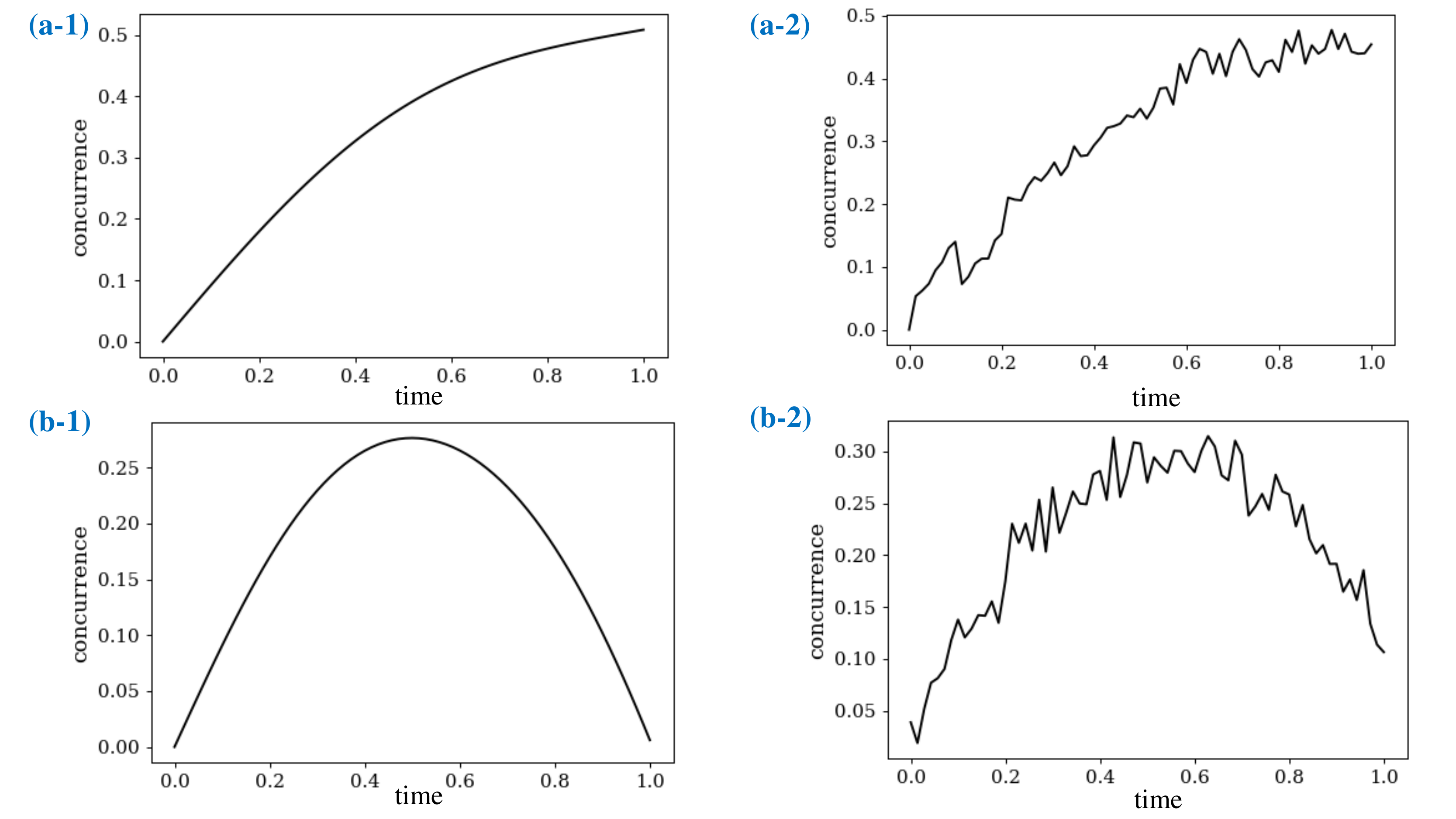}
\centering
\caption{The concurrence of the (a-1) state \eqref{initial2} under evolution \eref{evol} and (a-2) its simulated state reconstructed by  partial tomography process by $5$ observables, see Section \ref{robustness}. Here, $B=1$, $b=0.5$, $\lambda=1$, $\gamma=2\arccos\sqrt{3/4}$, $J=1$, and $\kappa=0.95$.
Plots of panel (b) show the same quantities for the state obtained once we change magnetic field and its inhomogenity as   $B=2.5$, and$b=0.25$.
Simulation results are obtained from ibmq-manila device on $2021-05-23$ with $8000$ shots.}
\label{conc2}
\end{figure}

The results are shown in figures \ref{fid2} and \ref{conc2}. Where in the first one we analysed the idea of robustness of X-states by applying the notion of fidelity.  Moreover, in the second figure the analytical results for  concurrence as a function of time is compared for several values of the magnetic field $B$ and the inhomogeneity  $b$ for fixed values of $\lambda$, $J$ and $\kappa$, with the one obtained by the actual IBM quantum device.\\

We end this section by emphasizing on the fact that the simplification proposed for simulation of the Heisienberg XYZ interaction is valid as long as the set of input states are restricted to the special set of classically correlated states. Once simulation of the time evolution of a general input state is desired, we should apply the common approach of  finding the gate decomposition of the unitary operator. Such a gate decomposition for a general $SU(4)$ unitary operator in $X$ shape as well as the special operator of Heisenberg system is provided in Appendix \ref{app}.



\section{Discussion}\label{disc}
\noindent We have extended the simulation study of Werner states \cite{PM19} and Bell-Diagonal-States of \cite{G19} to the larger set of X-states. The space of all BDSs is in one to one correspondence with points of  a tetrahedron and it has been shown in \cite{G19} that the agreement between the theoretical results on correlation properties of these states and the ones obtained by the Qiskit software and the IBM quantum device, while being satisfactory for most of the parameter space, deteriorates as we move towards the edges of the tetrahedron. This indicates that the noise model implemented in the Qiskit software does not fully represent the actual noise in the IBM quantum device. For each point of the tetrahedron, an X-state has two extra parameters denoted by $\theta$ and $\phi$ and shown in figure \ref{e1}.
 To see how much this agreement is kept intact, we have now explored the
 analogous problem for the X-states which are a mixture of equi-entangled basis.  These results are shown in figure \ref{t1} and they confirm that the more entangled states are supposed to be the more decrease in their fidelity happens.\\

A very serious issue related to the simulation of states covering a wide range of parameters concerns the run-time of simulations. One solution to overcome this problem is to find the symmetries preserved under environmental noises and gate imperfections.
When it comes to X-states, we have shown their shape is almost robust under these noises. This we have studied through substitution of state tomography by a partial tomography process and results are presented in figures \ref{fid1} and \ref{fid2}. Through this fact, one can simulate many states even by publicly available accounts on IBM quantum computer.\\

\noindent Furthermore, we have cast this study into a physically interesting model, namely the Heisenberg XYZ spin system and have performed the same study as before for the dynamical density matrix which results from this physical model, once the initial state is a classically correlated X-state. The results of this part are shown in figures \ref{conc} and \ref{conc2}.
Moreover, while we have already shown how to simplify simulation of the Heisenberg model
by taking the initial state to be a classically correlated state,  one might be interested in simulating the evolution of a general
state undergoing such an interaction rather than the classical one.
To model the time evolution of a general two-qubit state on a  quantum computer one needs the gate decomposition
of a general XYZ system. Such a decomposition is presented in Appendix \ref{app}.\\

Finally, it is worth mentioning the approach applied here for simulation of a general X-state is to implement the state \eqref{classic} on two qubits and then change this state to an X-state. Thus, we first encode the eigenvalues of the desired state in the classical probabilities $p_{i,j}$ and then transform the product basis $\ket{i,j}$ into eigenbasis of X-states by applying a proper unitary transformation. This approach can be applied to also simulate a general two-qubit state. The only thing one needs to do in this connection is to find the unitary evolution that transforms the product basis into the eigenbasis of the state whose simulation is desired. Note that the gate decomposition for a general unitary evolution is already known \cite{VW04}.\\

{\it Acknowledgement}--- This research was partially supported by the grant no. G98024071 from Iran National Science Foundation.
Financial support by Narodowe Centrum Nauki under the Grant No. DEC-2015/18/A/ST2/00274 is gratefully acknowledged.

\appendix
\section{Analytical Solution of Eq. \eqref{spec-shape}}\label{parameters}
Here, we present how to set the parameters of the circuit \ref{c1}
to achieve an assumed X-state. 
Given an X-state, six parameters $(a,b,c,d,w,z)$ in 
Eq.~\eqref{spec-shape} are known, hence one needs to invert these
equations to adjust the circuit parameters. 
It should be mentioned that setting the three parameters 
$(\alpha, \beta,\gamma)$ in the first block of Fig.~\ref{c1} for a given 
four dimensional probability vector $\{p_{ij}\}$ is already known
\cite{G19}.  What remains is to find the probabilities along with 
$(\theta,\phi)$ based on the parameters of an X-state.
Note that equations of $(a,d,w)$ in \eqref{spec-shape}
are independent of the remaining three parameters. 
Consider an X-state with $a\neq d$. In this case, we see through
Eq.~\eqref{spec-shape}
\begin{equation}
\theta=\frac12\arctan(\frac{2w}{a-d}).
\end{equation}
Having $\theta$ one gets the following probabilities 
\begin{equation}
p_{00}=\frac{a\cos^2\theta-d\sin^2\theta}{\cos2\theta},
\quad\quad
p_{10}=\frac{-a\sin^2\theta+d\cos^2\theta}{\cos2\theta}.
\end{equation}
It remains to find the case with $a=d$. In this situation, 
we should set $\theta=\pi/4$, $p_{00}=a+w$, and $p_{10}=a-w$. 
The remaining three parameters $(\phi,p_{01},p_{11} )$ are obtained 
through respective similar equations by substituting 
$a\leftrightarrow b$, $c\leftrightarrow d$, and $w\leftrightarrow z$.
\section{Simulation of Heisenberg XYZ system}\label{app}
To simulate evolution of a general initial state undergoing Heisenberg XYZ Hamiltonian, one may notice the resultant state in this case
is not necessarily an X-state. Thus a generalization of the approach mentioned in this paper is required. Indeed, to this end, we need to simulate the  unitary operator corresponding to the dynamics of the Heisenberg model irrespective of the initial state.
The most general operator in $U(4)$ which is in X-shape, has eight real parameters. If we drop an overall phase and restrict ourselves to X-operators in the group $SU(4)$, we are left with seven parameters. We parameterize such operators as follows:
\begin{equation}
U_X=
\begin{pmatrix}
e^{\frac{1}{2} i (a_1-x)} \cos t_1 & 0 & 0 & -e^{\frac{1}{2} i (b_1-x)} \sin t_1 \\
0 & -i e^{\frac{1}{2} i (a_2+x)} \sin t_2 & i e^{\frac{1}{2} i (b_2+x)} \cos t_2 & 0 \\
0 & i e^{-\frac{1}{2} i (b_2-x)} \cos t_2 & i e^{-\frac{1}{2} i (a_2-x)} \sin t_2 & 0 \\
e^{-\frac{1}{2} i (b_1+x)} \sin t_1 & 0 & 0 & e^{-\frac{1}{2} i (a_1+x)} \cos t_1 \\
\end{pmatrix}.
\end{equation}
The gate decomposition of such special unitary evolution in X-shape
is given by
\begin{equation}\label{u-gate}
U_X=\Big(R_z(s_1)\otimes R_z(s_2)\Big)C_1
\Big(R_y(t_1+t_2)\otimes W(x)\Big)C_2
\Big(R_y(t_1-t_2)\otimes I\Big)C_1\Big(R_z(s_3)\otimes R_z(s_4)\Big),
\end{equation}
where $W(x)=\diag\{\e^{-ix},i\e^{ix}\}$, and the two-qubit operators
$C_1$ and $C_2$ are CNOT gates whose control bits are respectively the first and the second qubits. Here  $R_y$
and $R_z$ are $SU(2)$ rotations around $Y$ and $Z$ axes. The parameters $s_i$ for
$i\in\{1,\dots,4\}$ are given by:
\begin{eqnarray}
s_1&=&-\frac14 (a_1 + b_1 + a_2 + b_2),\quad
s_2=-\frac14 (a_1 + b_1 - a_2 - b_2),\nonumber\\
s_3&=&-\frac14 (a_2 - b_2 + a_1 - b_1),\quad
s_4=-\frac14 (a_1 - b_1 - a_2 + b_2).
\end{eqnarray}
In view of (\ref{evol}), the gate decomposition of the unitary evolution for Heisenberg system is much simpler than above. In fact we see from (\ref{evol}) that
\be
x=J_zt, \ \ \  a_1=b_2=0,\ \ \  a_2=b_1=\pi,\ \ \  t_1=J\kappa t,\ \ \  t_2=Jt+\frac{\pi}{2}.
\ee
This means that the gate decomposition of the Heisenberg evolution operator (with anisotropic couplings and inhomogeneous magnetic field) is given by
\begin{equation}\label{u-gate}
U_X=\Big(\sigma_z\otimes I\Big)C_1
\Big(R_y(t_1+t_2)\otimes W(x)\Big)C_2
\Big(R_y(t_1-t_2)\otimes I\Big)C_1\Big(I\otimes \sigma_z\Big).
\end{equation}

\end{document}